\documentclass[prl,twocolumn,nofootinbib]{revtex4-1}
\usepackage[dvipsnames]{xcolor}
\usepackage{color}
\usepackage{graphicx}
\usepackage{color}
\usepackage{dcolumn}
\usepackage{bm}
\usepackage{amssymb}
\usepackage[bottom]{footmisc}
\usepackage{footnote}
\usepackage{lipsum}
\usepackage{gensymb}
\usepackage{wrapfig}
\usepackage{amsmath}
\usepackage{sidecap}
\usepackage[colorlinks,linkcolor=Blue,urlcolor=Blue,citecolor=Blue]{hyperref}
\renewcommand{\thefootnote}{\fnsymbol{footnote}}

\newcommand\blfootnote[1]{%
\begingroup
\renewcommand\thefootnote{}\footnote{#1}%
\addtocounter{footnote}{-1}%
\endgroup}

\begin{document}

\title{Ferromagnetic–antiferromagnetic coexisting ground states and exchange bias effects in $\bf{MnBi_4Te_7}$ and $\bf{MnBi_6Te_{10}}$}

\author{Xiaolong Xu,$^{1,2,3,\dagger}$ Shiqi Yang,$^{1,4,\dagger}$ Huan Wang,$^{5}$ Roger Guzman,$^{6,7}$ Yaozheng Zhu,$^{1}$ Yuxuan Peng,$^{1}$ Zhihao Zang,$^{1}$ Ming Xi,$^{5}$ Shangjie Tian,$^{5}$ Yanping Li,$^{1}$ Hechang Lei,$^{5}$ Zhaochu Luo,$^{1}$ Jinbo Yang,$^{1}$ Tianlong Xia,$^{5,\star}$ Wu Zhou,$^{6,7,\star}$ Yuan Huang,$^{3,\star}$ and Yu Ye,$^{1,2,8,\star}$}

\affiliation{$^{1}$State Key Laboratory for Mesoscopic Physics and Frontiers Science Center for Nano-optoelectronics, School of Physics, Peking University, Beijing 100871, China}
\affiliation{$^{2}$Collaborative Innovation Center of Quantum Matter, Beijing 100871, China}
\affiliation{$^{3}$Advanced Research Institute of Multidisciplinary Science, Beijing Institute of Technology, Beijing 100081, China}
\affiliation{$^{4}$Academy for Advanced Interdisciplinary Studies, Peking University, Beijing 100871, China}
\affiliation{$^{5}$Department of Physics, Renmin University of China, Beijing 100872, China}
\affiliation{$^{6}$School of Physical Sciences, University of Chinese Academy of Sciences, Beijing 100049, China}
\affiliation{$^{7}$CAS Centre for Excellence in Topological Quantum Computation, University of Chinese Academy of Sciences, Beijing 100049, China}
\affiliation{$^{8}$Yangtze Delta Institute of Optoelectronics, Peking University, Nantong 226010 Jiangsu, China}


\maketitle
\blfootnote{$^\dagger$ These authors contribute equally.}
\blfootnote{$^\star$Corresponding to: tlxia@ruc.edu.cn, wuzhou@ucas.ac.cn, yhuang@bit.edu.cn, and ye\_yu@pku.edu.cn}


\textbf{Natural superlattice structures $\bf{(MnBi_2Te_4)(Bi_2Te_3)_{\emph{n}}(\emph{n} = 1, 2,...)}$, in which magnetic $\bf{MnBi_2Te_4}$ layers are separated by nonmagnetic $\bf{Bi_2Te_3}$ layers, hold band topology, magnetism and reduced interlayer coupling, providing a promising platform for the realization of exotic topological quantum states. However, their magnetism in the two-dimensional limit, which is crucial for further exploration of quantum phenomena, remains elusive. Here, complex ferromagnetic (FM)–antiferromagnetic (AFM) coexisting ground states that persist up to the 2-septuple layers (SLs) limit are observed and comprehensively investigated in $\bf{MnBi_4Te_7}$ ($\emph{n}$ = 1) and $\bf{MnBi_6Te_{10}}$ ($\emph{n}$ = 2). The ubiquitous Mn-Bi site mixing modifies or even changes the sign of the subtle inter-SL magnetic interactions, yielding a spatially inhomogeneous interlayer coupling. Further, a tunable exchange bias effect is observed in $\bf{(MnBi_2Te_4)(Bi_2Te_3)_{\emph{n}}(\emph{n} = 1, 2)}$, arising from the coupling between the FM and AFM components in the ground state. Our work highlights a new approach toward the fine-tuning of magnetism and paves the way for further study of quantum phenomena in $\bf{(MnBi_2Te_4)(Bi_2Te_3)_{\emph{n}}(\emph{n} = 1, 2,...)}$ as well as their magnetic applications.}

\bigskip

The interplay between magnetism and topology inaugurates a new horizon in exploring exotic quantum phenomena, such as the quantum anomalous Hall effect (QAHE), axion insulators and magnetic Weyl semimetals\cite{N40,N1,N2,N3,N4,N5}. Recently, $\rm{MnBi_2Te_4}$ was discovered to be an intrinsic stoichiometric antiferromagnetic (AFM) topological insulator (TI)\cite{N7,N8,N10,N6,N9,N11,N39}. The intertwined band topology and magnetic order in A-type AFM $\rm{MnBi_2Te_4}$ (MBT) pose great challenges to the realization of its exotic topological physics and subsequent devices, because either the layer number of the material needs to be strictly controlled, or a high magnetic field ($\sim$ 6 to 8 T) is required \cite{N6,N9,N39,N11}. Therefore, there is an urgent need to develop a magnetic topological insulator of the MBT family that is magnetically insensitive to the number of layers and has a small saturation field, in which engineering of the interlayer magnetic interaction is the key. Now, the natural superlattice structures $\rm{(MnBi_2Te_4)(Bi_2Te_3)_\emph{n}(\emph{n} = 1, 2,...)}$ provides an opportunity to modify the interlayer exchange coupling using non-magnetic $\rm{Bi_2Te_3}$ (BT) quintuple layer (QL) as spacer layers\cite{N37,N12,N13,N41,N14,N15}. With the increase of n, the interlayer AFM coupling gradually weakens, while the global spin-orbit coupling strength gradually increases as the increase of Bi content\cite{N12,N13,N15,N14,N41,N44}. For $\rm{MnBi_4Te_7}$ ($\emph{n}$ = 1) and $\rm{MnBi_6Te_{10}}$ ($\emph{n}$ = 2), neutron scattering experiments and theoretical calculations show that they possess weak interlayer magnetic coupling, but still maintain the A-type AFM structure\cite{N14,N15,N16,N17,N18}, that is, in each $\rm{MnBi_2Te_4}$ septuple layer (SL), the Mn magnetic moments are ferromagnetically aligned, while the adjacent SLs are antiferromagnetically aligned. Although tremendous efforts have been devoted to studying the magnetic properties of the $\rm{MnBi_4Te_7}$ and $\rm{MnBi_6Te_{10}}$, most studies have only focused on their bulk phase\cite{N12,N13,N14,N41, N15,N19,N35,N42,N46}. However, the magnetic property and topological phase in MBT family often exhibit intricate thickness dependence, posing a significant influence for the realization of exotic topological physics\cite{N6,N9,N11,N21,N39,N22,N20}. Therefore, determining the magnetism of $\rm{(MnBi_2Te_4)(Bi_2Te_3)_\emph{n}(\emph{n} = 1, 2)}$ at their 2D limit is crucial, but has so far remained elusive.

In this work, we systematically investigate the magnetic properties of $\rm{MnBi_4Te_7}$ and $\rm{MnBi_6Te_{10}}$ thin flakes down to 1-SL by employing polar reflective magnetic circular dichroism (RMCD) spectroscopy. We demonstrate that the odd-even-layer oscillation of compensated and uncompensated AFM states, a common effect in ideal A-type AFM materials\cite{N11,N20,N23,N24}, vanishes in atomically thin $\rm{MnBi_4Te_7}$ and $\rm{MnBi_6Te_{10}}$. In all the measured samples above 1 SL, in addition to the AFM spin-flip transition, a significant FM hysteresis loop centered at zero field is observed. We reveal that the peculiar hysteresis loop arises from the complex FM-AFM coexisting ground state. Atomic-resolution electron energy loss spectroscopy (EELS) mapping as well as single-crystal X-ray diffraction (SC-XRD) reveal the ubiquitous Mn-Bi site mixing in the crystals. The spins of the randomly distributed $\rm{Mn_{Bi}}$ antisite defects in each SL are antiferromagnetically coupled to the spins of the main Mn layer\cite{N19,N25,N18}, that further modify or even change the sign of the subtle inter-SL magnetic interactions, yielding a spatially inhomogeneous interlayer coupling. As a result, the energy gained through the formation of magnetic domains compensates for the energy cost in the appearance of domain walls, yielding a complex FM-AFM coexisting ground state. Further, a tunable exchange bias effect is observed in $\rm{MnBi_4Te_7}$ and $\rm{MnBi_6Te_{10}}$, arising from the coupling between the FM and AFM components in the ground state. The direction of this exchange bias can be easily tuned by the historical polarization field and does not require warming and field cooling processes. Our results reveal the nontrivial magnetic ground states in $\rm{MnBi_4Te_7}$ and $\rm{MnBi_6Te_{10}}$, keeping the promise for further investigation of exotic topological quantum states.

\begin{figure*}[tb]
	\includegraphics[width=2\columnwidth]{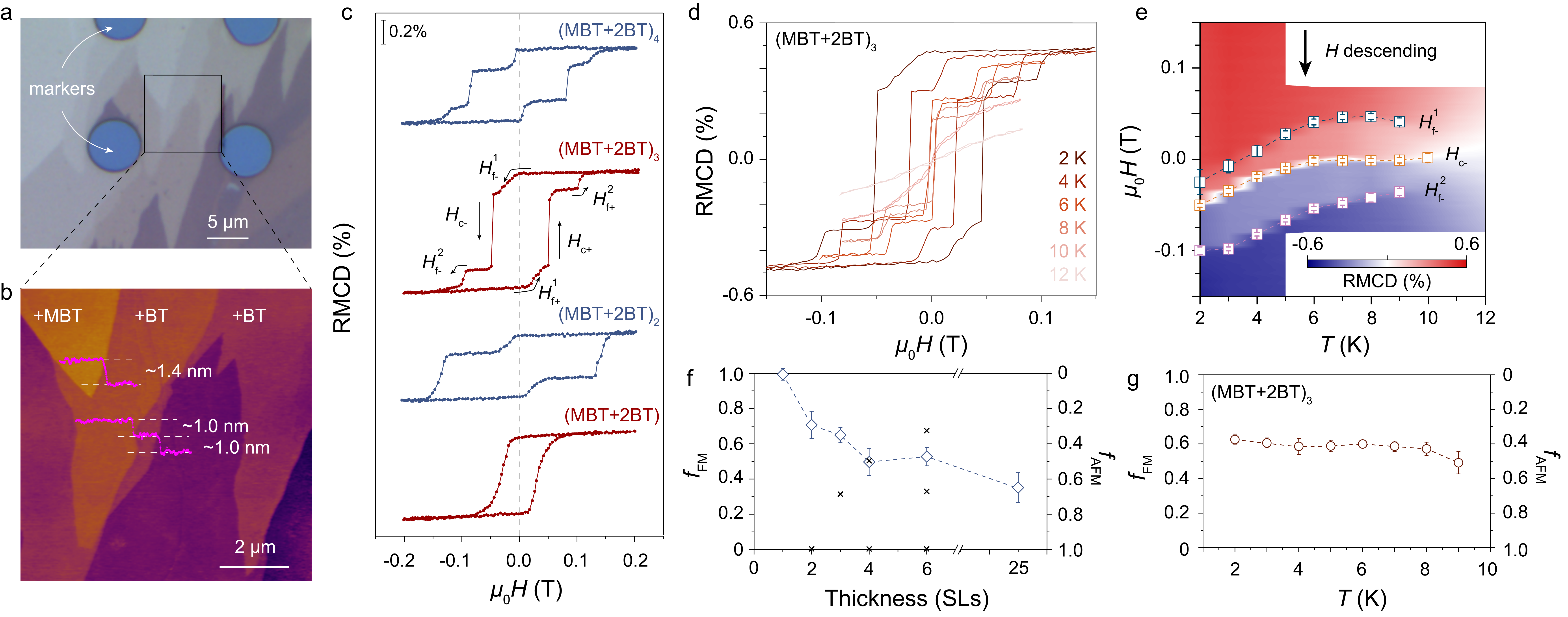}
	\caption{\label{F1}\textbf{FM-AFM coexisting magnetic ground state in $\bf{MnBi_6Te_{10}}$.} \textbf{a}, A typical optical image of the exfoliated $\rm{MnBi_6Te_{10}}$ sample with different thicknesses and terminations. \textbf{b}, Atomic force microscopy height image and height line profiles measured in the area marked by the black box in \textbf{a}, showing a thicknesses of $\sim$1.4 nm for $\rm{MnBi_2Te_4}$ SL and $\sim$1.0 nm for $\rm{Bi_2Te_3}$ QL. \textbf{c}, RMCD signals \textit{versus} $\mu_0H$ for 1 to  4-SL (terminated by 2-QL protection layer) $\rm{MnBi_6Te_{10}}$ flakes at 2 K. \textbf{d}, RMCD sweeps for the  3-SL $\rm{MnBi_6Te_{10}}$ flake at a temperature range that passes through its $T_{\rm{N}}$. \textbf{e}, 2D plot of RMCD signal as functions of temperature and magnetic field (descending sweep) for the  3-SL $\rm{MnBi_6Te_{10}}$ flake. Two AFM spin-flip fields ($H_{\rm{f-}}^1$ and $H_{\rm{f-}}^2$) and one FM spin-flip field ($H_{\rm{c-}}$) are extracted and superimposed on the plot. \textbf{f,g}, FM ($f_{\rm{FM}}$, left axis) and AFM ($f_{\rm{AFM}}$, right axis) component proportions as functions of thickness (\textbf{f}) and temperature (\textbf{g}). Cross marks in \textbf{f} indicate the expected values for the single-domain case. The vertical error bars are estimated from the uncertainty of the measured signal.}
\end{figure*}

\bigskip

\noindent
\textbf{FM-AFM coexisting magnetic ground states in $\bf{MnBi_4Te_7}$ and $\bf{MnBi_6Te_{10}}$ samples.} High-quality $\rm{MnBi_4Te_7}$ and $\rm{MnBi_6Te_{10}}$ bulk crystals are grown by flux method\cite{N33} and confirmed by XRD results (Fig. S1). The field-cooled (FC) and zero-field-cooled (ZFC) magnetic susceptibilities of $H\parallel c$ ($\chi^c$) of $\rm{MnBi_4Te_7}$ and $\rm{MnBi_6Te_{10}}$ crystals show that their long-range AFM orders occur at Néel temperature ($T_{\rm{N}}$) of 12.1 K and 10.9 K, respectively (Fig. S2). Compared with $\rm{MnBi_2Te_4}$\cite{N16,N17}($T_{\rm{N}}$ of $\sim$ 24.5 K), the reduction of $T_{\rm{N}}$ in $\rm{MnBi_4Te_7}$ and $\rm{MnBi_6Te_{10}}$ manifests the weakened interlayer coupling. Scrutiny of the $M-H$ curve at 2 K under $H\parallel c$, we find that $\rm{MnBi_4Te_7}$ ($\rm{MnBi_6Te_{10}}$) undergoes a spin transition at a very low magnetic field and quickly enters the forced FM state with a field of about 0.25 T (0.21 T). This is in sharp contrast to $\rm{MnBi_2Te_4}$, where the spin-flop transition occurs at about 3.5 T and its magnetic moment eventually saturates under external magnetic fields larger than 8 T\cite{N6,N9,N11,N20} (Fig. S3a). The much smaller saturation fields again indicate a significantly reduced interlayer AFM exchange interactions in $\rm{MnBi_4Te_7}$ and $\rm{MnBi_6Te_{10}}$.

 \begin{figure}[tb]
	\includegraphics[width=1\columnwidth]{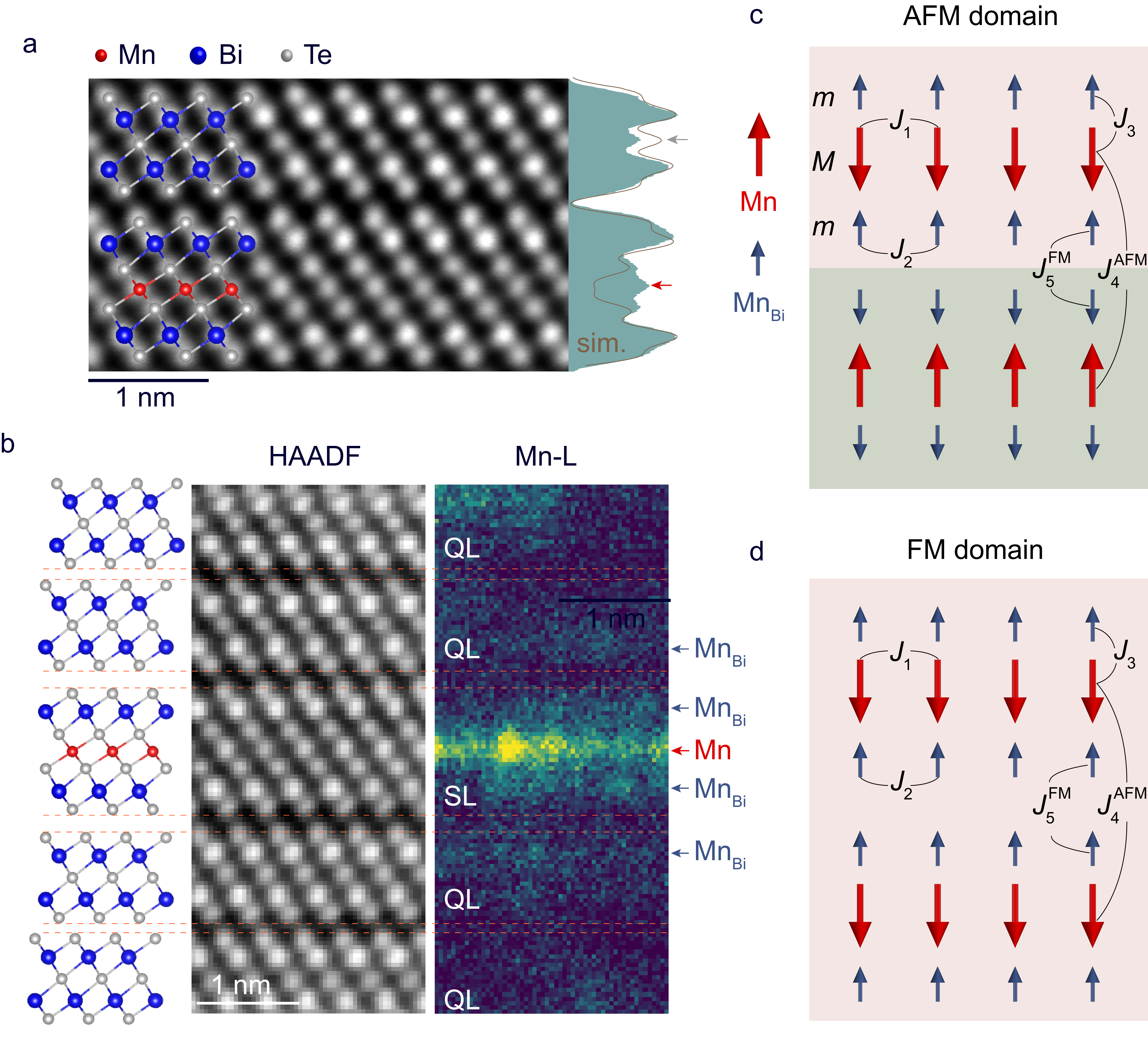}
	\caption{\label{F2}\textbf{STEM characterizations and an intuitive model of inter-SL FM and AFM orders in $\bf{MnBi_4Te_7}$ and $\bf{MnBi_6Te_{10}}$}. \textbf{a}, Atomic-resolution HAADF image of the cross-section of the $\rm{MnBi_6Te_{10}}$ crystal along the [100] direction. Experimental (shaded area) and simulated (grey curve) integral HAADF intensity profiles along the $c$-axis show clear discrepancy, showing $\rm{Bi_{Mn}}$ antisite defects in SL (red arrow) and $\rm{Mn_{Te}}$ in QL (grey arrow). \textbf{b}, Atomic structure and HAADF image of the $\rm{MnBi_6Te_{10}}$ crystal with the corresponding EEELs mapping of the Mn element ($\rm{L_{2,3}}$ edge). Clear Mn signals are present in the Bi layers of SL and QL. The orange dashed lines indicate the van der Waals gaps. \textbf{c,d}, Illustrations of the inter-SL FM and AFM regions in $\rm{MnBi_4Te_7}$ and $\rm{MnBi_6Te_{10}}$, in which the dominant magnetic exchange interactions are labelled.}
\end{figure}

To explore how the weakened interlayer exchange interaction affects the magnetic ground order, we investigated their layer-number-dependent magnetism of $\rm{MnBi_4Te_7}$ and $\rm{MnBi_6Te_{10}}$ using the RMCD spectroscopy. Atomically thin flakes down to 1 SL are mechanically exfoliated from their bulk crystals onto gold substrates using the standard Scotch tape method\cite{N27}, and then protected with a layer of polymethyl methacrylate (PMMA) (Fig. \ref{F1}a). The number of layers is confirmed by atomic force microscopy characterizations (see Fig. S4 for $\rm{MnBi_4Te_7}$ samples and Fig. S5 for $\rm{MnBi_6Te_{10}}$ samples). Typical height line profiles of the stepped $\rm{MnBi_6Te_{10}}$ flakes indicate SL and QL thicknesses of approximately 1.4 and 1.0 nm, respectively (Fig. \ref{F1}b), consistent with previous reports \cite{N22,N28}. The layer-number-dependent magnetic behavior in $\rm{MnBi_4Te_7}$ and $\rm{MnBi_6Te_{10}}$ is very similar, so here we show 1-4 SLs $\rm{MnBi_6Te_{10}}$ as examples, and provide the full set of other samples in Figs. S6 to S9. 1-SL $\rm{MnBi_6Te_{10}}$ shows a distinct RMCD signal at zero field and a clear hysteresis loop (Fig. \ref{F1}c), confirming its FM nature. With increasing temperature, the hysteresis loop shrinks and disappears at 10 K, indicating an FM to paramagnetic phase transition (Fig. S8a). Compared with the intrinsic 1-SL $\rm{MnBi_2Te_4}$ ($T_{\rm{C}}$ = 14.5 K)\cite{N20}, the decreased $T_{\rm{C}}$ may be due to the increased $\rm{Bi_{Mn}}$ sites, as the intralayer exchange coupling decreases with the increase of the average distance between the Mn sites. Surprisingly, with increasing thickness, no odd-even layer-number oscillation occurs, which is a typical feature of ideal A-type AFM materials. Instead, both FM and AFM signals are present in all the measured samples with layer number $N$\textgreater1 (Fig. \ref{F1}c). In the 2-SL samples, one AFM and one FM spin-flip transitions occur during the magnetic reversal process, signifying the coexistence of ↑↑ and ↑↓ domains. With increasing thickness, the magnetic reversal curve evolves into two AFM and one FM spin-flip transitions with different magnetic moment amplitudes.

As the temperature increases, the value of the spin-flip field $H_{\rm{f-}}^1$ in the 3-SL sample changes from negative to positive, and then slowly decreases to approach zero (Fig. \ref{F1}d, e). A positive value of $H_{\rm{f-}}^1$ signifies the AFM nature of this spin-flip process, since interlayer AFM coupling prefers to make the magnetic moment in adjacent layers be antiparallel, while the Zeeman energy tends to keep the magnetic moment parallel to the magnetic field. Since the AFM interlayer coupling is stronger in the $\rm{MnBi_4Te_7}$ samples, $H_{\rm{f-}}^1$ usually occurs at larger positive values (see Figs. S6 and S9). Different coupling types and coupling strengths between AFM and FM components will result in distinct hysteresis behaviors. The FM spin-flip field $H_{\rm{c-}}$ of the $\rm{MnBi_6Te_{10}}$ samples increase monotonically from a negative value and approach zero with increasing temperature (Figs. \ref{F1}d, e and Fig. S8b). However, in some $\rm{MnBi_4Te_7}$ samples (Fig. S9d), $H_{\rm{c-}}$ first increases from negative to positive and then jumps back to negative with increasing temperature, signifying the correlated coupling between the FM and AFM components. Specifically, we build a five-layer macrospin model to interpret the hysteresis loops in the $\rm{MnBi_4Te_7}$ and $\rm{MnBi_6Te_{10}}$ samples (Fig. S10), which captures the main features. 

In real materials, the FM and AFM distributions and proportions can be very complex due to the spatial inhomogeneity. We denote the proportions of FM and AFM components by $f_{\rm{FM}} = \frac{\text{RMCD(FM flip height)}}{\text{RMCD(up saturation)}-\text{RMCD(down saturation)}}$ and $f_{\rm{AFM}} = 1-f_{\rm{FM}}$, respectively. As expected, $f_{\rm{AFM}}\sim0$ ($f_{\rm{FM}}\sim1$) in the 1-SL sample (Fig. \ref{F1}f). The $f_{\rm{AFM}}$ ($f_{\rm{FM}}$) of $\rm{MnBi_{6}Te_{10}}$ samples increases (decreases) as the number of layers increases. In most samples, the values of $f_{\rm{AFM}}$ do not coincide with those expected for single-domain case (cross marks in Fig. \ref{F1}f), indicating a multi-domain configuration within the laser spot size. The values of $f_{\rm{FM}}$ and $f_{\rm{AFM}}$ are nearly temperature independent over the entire measurement temperature range (Fig. \ref{F1}g). 

After confirming the FM-AFM coexistence ground states, the puzzling magnetism of bulk $\rm{MnBi_4Te_7}$ and $\rm{MnBi_6Te_{10}}$ can also be well explained. It is worth noting that at low temperatures, the $\rm{MnBi_4Te_7}$ and $\rm{MnBi_6Te_{10}}$ crystals exhibit non-zero magnetization at zero field (Fig. S3), and the magnetic reversal completes through three sluggish spin-flip transitions (marked by the arrows in Fig. S3)\cite{N7,N12,N13,N14,N16,N22,N41}. The non-zero magnetization at the zero field suggests that there may be some FM components in the AFM sate. The FM-AFM coexisting magnetic order is also confirmed by the bifurcations of the ZFC and FC curves at temperatures slightly below the Néel temperature (Fig. S2). The large difference in the values of $f_{\rm{AFM}}$ between thin flakes and bulk crystals suggests that the AFM-FM coexisting magnetic order is possibly a surface-related effect. The complex multi-domain structure in the bulk samples smears the distinct signatures of the FM-AFM coexisting ground state, resulting in sluggish changes in the magnetic moment, which also masks the potential applications of this unique magnetic order.

\begin{figure*}[!tb]
	\includegraphics[width=1.5\columnwidth]{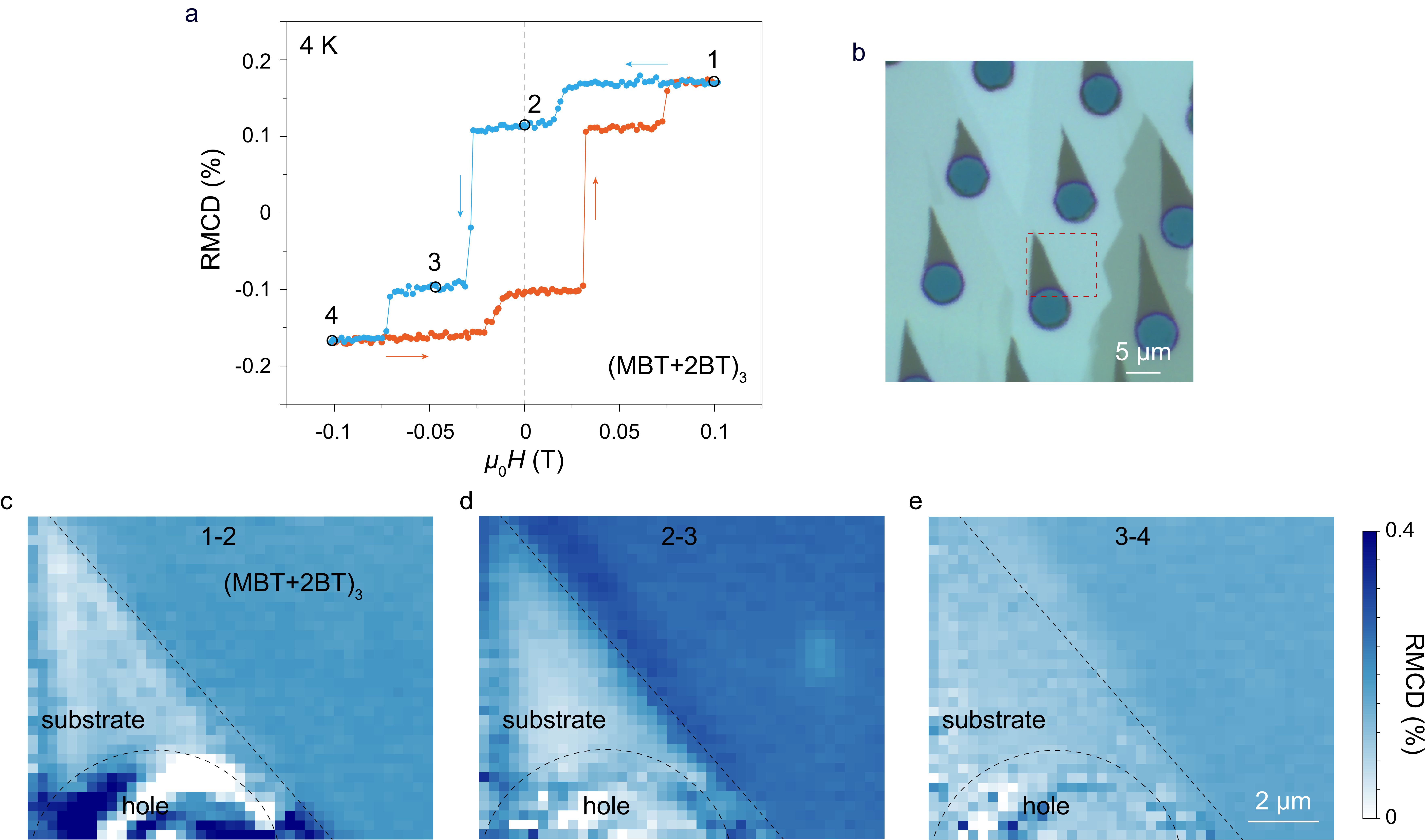}
	\caption{\label{F3}\textbf{Spatial RMCD mappings of $\bf{(MBT+2BT)_3}$ sample.} \textbf{a}, RMCD signal \textit{versus} external magnetic field of a 3-SL $\rm{MnBi_6Te_{10}}$ flake at 4 K. Four magnetic fields (labeled as 1 to 4) corresponding to four different spin configurations are selected for RMCD maping. \textbf{b}, Optical image of the stepped $\rm{MnBi_6Te_{10}}$ sample. The selected area for RMCD mapping is labeled by the dashed red box. \textbf{c-e}, Spatial distributions of the RMCD signal changes in two AFM spin-flip transitions (1 to 2 and 3 to 4) and one FM spin-flip transition (2 to 3). The homogeneous signals indicate the uniform FM-AFM coexistence under the experimental spatial resolution. The bare substrate and etched hole are marked by black dashed lines.}
\end{figure*}

\bigskip

\noindent
\textbf{Origin of FM-AFM coexisting magnetic ground state.} Due to the weak interlayer coupling, the energy difference between the AFM and FM spin configurations is very small \cite{N14,N15}, providing a breeding ground for the magnetism tuning\cite{N43,N45}. Cross-sectional atomic-resolution high-angle annular dark-field scanning transmission electron microscopy (HAADF-STEM) image along the [100] direction shows that two $\rm{Bi_2Te_3}$ layers are inserted between $\rm{MnBi_2Te_4}$ layers in $\rm{MnBi_6Te_{10}}$ (Figs. \ref{F2}a and b). However, the experimental and simulated integrated HAADF intensity profiles along the $c$-axis show clear discrepancy, showing $\rm{Bi_{Mn}}$ antisite defects in SL (red arrow in Fig. \ref{F2}a) and $\rm{Mn_{Te}}$ in QL (grey arrow in Fig. \ref{F2}a). In addition, atomic electron energy loss spectroscopy (EELS) mapping shows clear Mn signals in the Bi layers in SL and also QL (Fig. \ref{F2}b), demonstrating the existence of the $\rm{Mn_{Bi}}$ antisite defects (see Fig. S11 for $\rm{MnBi_4Te_7}$). 

Below ${T\rm{_N}}$, the spins of Mn atoms are ferromagnetically coupled ($-J_{1}$) within the SL, but antiferromagnetically coupled ($J_{4}$) with those of Mn atoms in the adjacent SL. The spins of $\rm{Mn_{Bi}}$ in SL are ferromagnetically coupled ($-J_{2}$) within the sublayer, but antiferromagnetically coupled ($J_{3}$) to the main Mn layer\cite{N25}, forming a ferrimagnetic SL configuration (Fig. \ref{F2}c). Flipping the spins of $\rm{Mn_{Bi}}$ in SL to align with those in the Mn layer requires a magnetic field above 50 T\cite{N25}, so all spins in SL flip together under a small magnetic field. This stable ferrimagnetic SL configuration results in a reduced measured saturation magnetic moment than the theoretically value\cite{N14} (Fig. S3), and determines that the magnetic moment change at low magnetic fields (from $-$0.2 T to 0.2 T) comes only from that of the main Mn layer. Considering that Mn atoms doped into the $\rm{Bi_2Te_3}$ always tend to be ferromagnetically coupled\cite{N29,N30,N31}, the spins of $\rm{Mn_{Bi}}$ are expected to be ferromagnetically coupled ($-J_{5}$) in the adjacent SLs and within QLs. The magnetic moment in the QL to flip to align with the magnetic filed requires a magnetic field of about 8 T and 6 T in $\rm{MnBi_4Te_7}$ and $\rm{MnBi_6Te_{10}}$\cite{N19}, respectively (marked by the arrows in Fig. S3). For simplification, we do not additionally consider the coupling in QL and accommodate it in $J_{5}$, which effectively strengthens the FM coupling between the $\rm{Mn_{Bi}}$ antisites in the two neighboring SLs. Then, the Hamiltonian of the 2-SL system can be expressed as:
\begin{equation}
\begin{split}
H & = -J_{1}\sum_{<ij>} \vec{M}_i^{1}  \cdot \vec{M}_j^{1} -J_{1}\sum_{<ij>}\vec{M}_i^{2} \cdot \vec{M}_j^{2} \\
&-J_{2}\sum_{<ij>}\vec{m}_i^{1} \cdot \vec{m}_j^{1} -J_{2}\sum_{<ij>}\vec{m}_i^{2}  \cdot \vec{m}_j^{2} \\
&+ J_{4}\sum_{<ij>}\vec{M}_i^{1} \cdot \vec{M}_j^{2} -J_{5}\sum_{<ij>}\vec{m}_i^{1} \cdot  \vec{m}_j^{2} \\
&+J_{3}\sum_{<ij>}\vec{M}_i^{1} \cdot  \vec{m}_j^{1} + J_{3} \sum_{<ij>}\vec{M}_i^{2} \cdot  \vec{m}_j^{2}
\end{split} 
\end{equation}
where $m_i$ and $M_i$ are the magnetic moments at each $\rm{Mn_{Bi}}$ and Mn lattice sites, superscripts 1 and 2 represent the SL layer indices, and the sum is over all the nearest-neighboring lattice sites. In the presence of $\rm{Mn_{Bi}}$ antisite defects, the energy difference per unit cell between the AFM and FM configurations (Figs. \ref{F2}c and d) can be expressed as:
\begin{equation}
\begin{aligned}
U_{\rm{AFM}}-U_{\rm{FM}}=2(J_5\sum_{<ij>}m_i^{1}m_j^{2}-J_4\sum_{<ij>}M_i^{1}M_j^{2})
\end{aligned}
\end{equation}
This energy difference increases with the amount of Mn-Bi site mixing (the numbers of $M_i$ sites decreases and the number of $m_i$ sites increases), eventually leading to a change in the sign of the interlayer magnetic coupling (from AFM to FM). The randomly distributed $\rm{Mn_{Bi}}$ antisites lead to a spatially inhomogeneous interlayer coupling. As a result, the FM-AFM coexisting ground states are expected to emerge when the energy gain from forming FM domains exceeds the energy cost from forming domain walls. In general, the lattice defect concentration of the MBT family is difficult to precisely control due to the narrow growth temperature window and the large size difference between Mn and Bi ions. Then, in the isostructural $\rm{MnSb_2Te_4}$ with a larger growth temperature window\cite{N32}, we observe the evolution of A-type AFM to FM-AFM coexistence and finally to the FM ground state with varying the Mn-Sb site-mixing concentration (Fig. S12).

\bigskip

\noindent
\textbf{Domain size and distribution characterizations.} To evaluate the domain sizes of the FM and AFM components, we characterize the magnetic spatial homogeneity by RMCD mapping. In the typical RMCD-$\mu_0H$ curve of a 3-SL $\rm{MnBi_6Te_{10}}$ sample (Fig. \ref{F3}a), we map the RMCD signals in a selected area (Fig. \ref{F3}b) under four selected magnetic fields (0.1 T, 0 T, $-$0.05 T, and $-$0.1 T, respectively) corresponding to four different spin configurations (see Fig. S13 for $\rm{MnBi_4Te_7}$ sample). 
Changes in the RMCD signal at the two AFM spin-flip transitions (Fig. \ref{F3}c and Fig. \ref{F3}e) and the FM spin-flip transition (Fig. \ref{F3}d) are uniform across the scanned sample area, indicating homogeneous AFM-FM coexistence at a spatial resolution limited by the laser spot size of $\sim$ 2 $\mu$m. The small domain size of the FM and AFM components is consistent with the large number of Mn-Bi site mixing characterized by single-crystal XRD (see Tables. \uppercase\expandafter{\romannumeral1} and \uppercase\expandafter{\romannumeral2}). 

\begin{figure*}[!tb]
	\includegraphics[width=2\columnwidth]{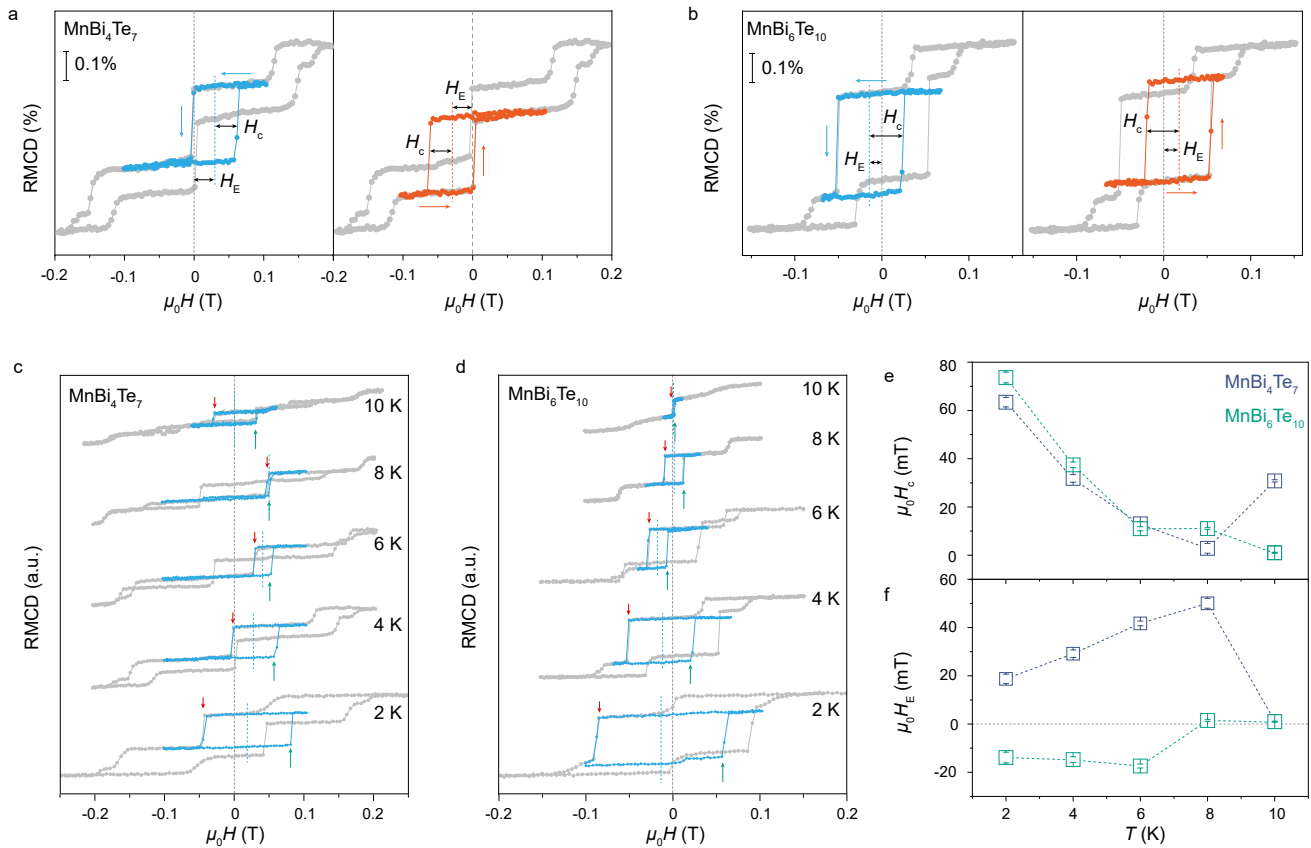}
	\caption{\label{F4} \textbf{Exchange bias effects of the FM component in $\bf{MnBi_4Te_7}$ and $\bf{MnBi_6Te_{10}}$ samples.} \textbf{a,b}, Exchange bias effects of the FM component in the $\rm{MnBi_4Te_7}$ and $\rm{MnBi_6Te_{10}}$ measured at 4 K. The direction of the exchange bias can be tuned by the historical polarization field. The exchange bias effect observed in $\rm{MnBi_6Te_{10}}$ is opposite to that of $\rm{MnBi_4Te_7}$. Colored dashed lines represent $H_{\text{E}}$. \textbf{c,d}, Temperature-dependent exchange bias effects of $\rm{MnBi_4Te_7}$ and $\rm{MnBi_6Te_{10}}$, with the full hysteresis loops being presented as the references. The red and green arrows represent the $H_{\rm{c-}}$ and $H_{\rm{c+}}$ for the biased hysteresis loop, respectively, and the blue dashed lines represent $H_{\text{E}}$. \textbf{e,f}, Evolution of $H_{\rm{c}}$ and $H_{\text{E}}$ with temperature in $\rm{MnBi_4Te_7}$ (blue squares) and $\rm{MnBi_6Te_{10}}$ (green squares).}
\end{figure*}

\bigskip

\noindent
\textbf{Tunable exchange bias from the FM-AFM coexisting ground state.} From the temperature-dependent hysteresis loops, we conclude that the coexisting FM and AFM components are correlated, which provides us with a unique system to study exchange bias effects. The large-field full hysteresis loops at 4 K (grey data in Figs. \ref{F4}a-d) are plotted as references for the minor hysteresis loops of the FM components. Historically polarized by a large positive saturation magnetic field, the minor hysteresis loop of the FM component shifts to the right side in $\rm{MnBi_4Te_7}$ sample (blue data in Fig. \ref{F4}a). The hysteresis behavior of the magnetic field descending sweeping is time reversal to that of the magnetic field ascending sweeping. Historically polarized by a large negative saturation magnetic field, then the minor hysteresis loop of the FM component shifts to the left, indicating a negative exchange bias (orange data in Figs. \ref{F4}b). The direction of this exchange bias can be easily tuned by the historical polarization field and does not require warming and field cooling processes. Moreover, this exchange bias is very stable under multiple back-and-forth magnetic field sweeps, with no training effect (Fig. S14). The exchange bias effect observed in $\rm{MnBi_6Te_{10}}$ is opposite to that of $\rm{MnBi_4Te_7}$, possibly due to the different magnetic structures at the FM/AFM interfaces because of the different interlayer coupling strengths (see discussion following Fig. S10). 

As the temperature increases, the coercive field $H_{\text{c}}$ for the biased hysteresis loop, defined as $1/2(H_{\text{c+}}-H_{\text{c-}})$, gradually shrinks and the exchange bias field $H_{\text{E}}$, defined as $1/2(H_{\text{c+}}+H_{\text{c-}})$, slightly increases (Figs. \ref{F4}c-f). The $H_{\text{c}}$ shows an abnormal increase at 10 K for $\rm{MnBi_4Te_7}$ and 8 K for $\rm{MnBi_6Te_{10}}$, respectively, accompanied by the vanishment of the exchange bias effect. In typical AFM and FM heterostructure systems, the exchange bias effect occurs only when $(K_\text{AFM}t_{\text{AFM}})/{J_{\text{int}}} \ge 1$, where $K_{\text{AFM}}$ is the anisotropy energy of the AFM component, $t_{\text{AFM}}$ is the thickness of the AFM component, and $J_{\text{int}}$ is the exchange coupling between the AFM and FM components. As the temperature increase, the $K_{\text{AFM}}$ in the MBT system decreases, and eventually, the AFM pinning layer flips collectively with the FM spins, resulting in the vanishment of the exchange bias and the abnormal increase in the coercive field. At high temperatures, the two AFM spin-flip transitions lead to identical RMCD signal changes, also confirming the collective flipping of the AFM and FM components during the FM spin-flip transition. 

\bigskip

In summary, we systematically study the magnetism of atomically thin $\rm{MnBi_4Te_7}$ and $\rm{MnBi_6Te_{10}}$ flakes in the parameter space of layer number, temperature and applied magnetic field using RMCD spectroscopy. The complex FM-AFM coexisting ground state is observed and verified. The weakened interlayer coupling and inhomogeneously distributed ubiquitous Mn-Bi site mixing have been attributed to result in such a unique magnetic ground state. A tunable exchange bias effect is observed in $\rm{MnBi_4Te_7}$ and $\rm{MnBi_6Te_{10}}$, arising from the coupling between the FM and AFM components. The demonstrated tunable exchange bias by introducing the coexistence of interlayer AFM and FM coupling provides design principles and materials for spintronic devices. Using sophisticated techniques, synthetic antiferromagnets composed of two or more ferromagnetic layers separated by non-magnetic spacers can be precisely prepared. Due to the weak interlayer exchange coupling in synthetic antiferromagnets, introducing spatial inhomogeneity in thickness of the spacer layer or disorders can therefore tune the interaction and form an FM-AFM coexisting ground state, allowing for precise manipulation of the exchange bias effect in this system\cite{N34}. By unraveling the puzzling magnetic states in $\rm{MnBi_4Te_7}$ and $\rm{MnBi_6Te_{10}}$ flakes, our findings pave the way for further investigation of quantum phenomena intertwined with the magnetic orders.

\section{Methods}
\noindent
$\textbf{Crystal growth and magnetic characterizations.}$ $\rm{(MnBi_2Te_4)(Bi_2Te_3)_{\emph{n}}(\emph{n} = 1, 2,...)}$ single crystals were grown by flux method \cite{N33}. Mn powder, Bi lump and Te lump were weighed with the ratio of Mn: Bi: Te = 1: 8: 13 (MnTe: $\rm{Bi_2Te_3}$ = 1: 4). The mixture was loaded into a corundum crucible sealed into a quartz tube. The tube was then placed into a furnace and heated to 1100 °C for 20 h to allow sufficient homogenization. After a rapid cooling to 600 °C at 5 °C/h, the mixture was cooled slowly to 585 °C (581 °C) at 0.5 °C/h for $\rm{MnBi_4Te_7}$ ($\rm{MnBi_6Te_{10}}$) and kept at this temperature for 2 days. Finally, the single crystals were obtained after centrifuging. The centimeter-scale plate-like $\rm{MnBi_4Te_7}$ and $\rm{MnBi_6Te_{10}}$ single crystals can be easily exfoliated. Magnetic measurements of $\rm{MnBi_4Te_7}$ and $\rm{MnBi_6Te_{10}}$ single crystals were measured by the vibrating sample magnetometer (VSM) option in a Quantum Design Physical Properties Measurement System (PPMS-9 T). The temperature-dependent magnetization measurements are described in detail in Supplementary Materials.
\bigskip

\noindent
$\textbf{Preparation of the ultra-thin samples.}$ The $\rm{MnBi_4Te_7}$ and $\rm{MnBi_6Te_{10}}$ flakes with different thicknesses were first mechanically exfoliated on a polydimethylsiloxane (PDMS) substrate by the Scotch tape method. The exfoliated samples on PDMS substrates were then dry-transferred onto 285 nm $\rm{SiO_2}$/Si substrates with evaporated gold films. Then, a layer of PMMA was spin-coated on the thin flakes for protection.
\bigskip

\noindent
$\textbf{AFM characterization.}$ The thickness of the ultra-thin samples wwas verified by the atomic force microscopy characterization using the Oxford Cypher S system in tapping mode. According to the height line profiles, the $\rm{MnBi_4Te_7}$ and $\rm{MnBi_6Te_{10}}$ were confirmed to possess an alternated lattice structure of BT (1 nm) + MBT (1.4 nm) and  BT (1 nm) + BT (1 nm) + MBT (1.4 nm). See more details in Supplementary Materials.
\bigskip

\noindent
$\textbf{RMCD measurements.}$ The RMCD measurements were performed based on the Attocube closed-cycle cryostat (attoDRY2100) down to 1.6 K and up to 9 T in the out-of-plane direction. The linearly polarized light of 633 nm HeNe laser was modulated between left and right circular polarization by a photoelastic modulator (PEM) and focused on the sample through a high numerical aperture (0.82) objective. The reflected light was detected by a photomultiplier tube (THORLABS PMT1001/M). The magnetic reversal under external magnetic field was detected by the RMCD signal determined by the ratio of the a.c. component of PEM at 50.052 kHz and the a.c. component of chopper at 779 Hz (dealt by a two-channel lock-in amplifier Zurich HF2LI). The errors of ratio of FM and AFM components are determined by the instability of the data acquired during RMCD measurements.
\bigskip

\noindent
$\textbf{STEM characterization.}$ Samples for cross-sectional investigations were prepared by standard lift-out procedures using an FEI Helios NanoLab G3 CX focused ion beam system. To minimize sidewall damage and make the samples sufficiently thin to be electronically transparent, final milling was carried out at a voltage of 5 kV and a fine milling at 2 kV. Aberration-corrected STEM imaging was performed using a Nion HERMES-100 operating at an acceleration voltage of 60 kV and a probe forming semi-angle of 32 mrad. HAADF images were acquired using an annular detector with a collection semi-angle of 75-210 mrad. EELS measurements were performed using a collection semi-angle of 75 mrad, an energy dispersion of 0.3 eV per channel, and a probe current of $\sim$20 pA. The Mn-$L$ (640 eV) and Te-$M$ (572 eV) absorption edges were integrated for elemental mapping after background subtraction. The original spectrum images were processed to reduce random noise using a principal component analysis (PCA) tool. HAADF image simulations were computed using the STEM\_CELL software simulation package matching the microscope experimental settings described above and using a supercell with a thickness $\sim$20 nm.

\section{\label{sec:level1}DATA AVAILABILITY}
The data that support the findings of this study will be available at an open-access repository with a doi link, when accepted for publishing.

\section{\label{sec:level3}ACKNOWLEDGEMENT}
This work was supported by the National Key R\&D Program of China (Grants No. 2018YFA0306900, No. 2017YFA0206301, 2019YFA0308602, No. 2019YFA0308000, and 2018YFA0305800), the National Natural Science Foundation of China (Grants No. 62022089, No. 12074425, and No. 11874422), Strategic Priority Research Program (B) of the Chinese Academy of Sciences (Grant No. XDB33000000), Beijing Natural Science Foundation (Grant No. JQ21018 and BJJWZYJH01201914430039), and the fundamental Research Funds for the Central Universities (E1E40209).

\section{Author contributions}
Y.Y., S.Y., and X.X. conceived the project, designed the experiments, analyzed the results and wrote the manuscript.  S.Y. and X.X. conducted the RMCD measurements. H.W. and T.X. grew the $\rm{MnBi_4Te_7}$ and $\rm{MnBi_6Te_{10}}$ bulk crystals. M.X., S.T., and H.L. grew the $\rm{MnSb_2Te_4}$ bulk crystal. Y.H. prepared the few-layer samples. Y.P. and J.Y. performed the magnetic characterizations of the bulk crystals. R.G. performed the STEM characteristics under the supervision of W.Z. Y.Z. and Z.L. helped with results analysis. All authors discussed the results and contributed to the manuscript.

\section{ADDITIONAL INFORMATION}
Competing interests: The authors declare no competing financial interests.

\bibliography{main.bib}
\bibliographystyle{naturemag}

\end{document}